
\input harvmac

\def\MPL#1{Mod.~Phys.~Lett. {\bf A{#1}}}

\def\PL#1{Phys.~Lett. {\bf B{#1}}}

\def\PRD#1{Phys.~Rev. {\bf D{#1}}}

\def\PRL#1{Phys.~Rev.~Lett. {\bf {#1}}}
\def\PTP#1{Prog.~Theor.~Phys. {\bf {#1}}}


%
%
\centerline{}
\vskip0.1cm
\Title{\vbox{\baselineskip12pt\hbox{}}}
{\vbox{\centerline{The Oblique Parameters In Electroweak Theory}
       \vskip2pt
       \centerline{With Two Massless Higgs Doublets}}}


\centerline{Kazunori Takenaga\footnote{$^\dagger$}
{E-mail address: f77498a@kyu-cc.cc.kyushu-u.ac.jp}}
\bigskip
\centerline{Department of Physics, Kyushu University}
\centerline{Fukuoka, 812 JAPAN}
\vskip 3cm
The oblique parameters $S$ and $T$ are studied in electroweak theory with two
massless Higgs doublets.
The effect of quadratic dependence on the charged Higgs mass in the parameter
$T$
could be cancelled.
The constraint on the charged Higgs mass from
the parameter $T$ is not so stringent as it was expected.
The parameter $S$ can be
both positive and negative but it can't take large negative value.
We search allowed scalar mass regions for both $S$ and
$T$ in the case of two different reference points.
\Date{12/93}



%
\newsec{Introduction}
The precise electroweak measurements at LEP have tested the standard model,
and they have given constraint on the top quark mass $M_t\simeq
131^{+47}_{-28}$ GeV
\ref\lang{P. Langacker, Lectures presented at TASI-92, Boulder, June 1992.}.
Now the electroweak measurements reach a sensitivity at which they are
probing
effects of new physics beyond the standard model or constraining it.
If new heavy particles exist and couple weakly only to light fermions, then
their
effect on low-energy electroweak phenomena appears through
contributions to gauge-boson self-energies. This type of radiative correction
is
called an oblique correction. The oblique corrections are summarized by three
parameters $S$, $T$ and $U$
 \ref\pt{M. Peskin and T. Takeuchi, \PRL{65}, (1990), 964. \PRD{46}, (1992),
381.}.
These
parameters are defined in terms of the gauge-boson self-energies. The $S$,
$T$
and $U$ are measures of the size of the radiative corrections arising from
new
physics beyond the standard model. The neutral current and many low-energy
observables depend on
$S$ and $T$, and the only measured weak interaction observable which depends
on $U$
is
$M_W$.
$U$ is in general suppressed compared with $T$ by a factor $M_Z^2/M_{new}^2$,
where $M_{new}$ is a typical mass scale of new physics. $U$ is often
predicted to be
very small. So one usually assumes $U\simeq 0$ and restricts
$S$ and
$T$ by the neutral current, $M_W$ and the other low-energy observables.
Experimental limits on $S$ and $T$ differ depending on the values of $M_t$
and
$M_H$ chosen as the reference standard model.
It seems that negative values of both $S$ and $T$ are favored \pt\ though
experimental and theoretical uncertainties in atomic parity violation are
significant
for $S$ \ref\mr{W. J. Marciano and J. L. Rosner, \PRL{65}, (1990), 2963}.
\nref\dr{M. J. Dugan and L. Randall, \PL{264}, (1991), 154.}
\nref\gt{E. Gate and J. Terning, \PRL{67}, (1991), 1840.}
\nref\ls{M. A. Luty and R. Sundrum, Preprint, LBL-32893, UCB-PTH-92/34.}
\nref\ILY{ T. Imami, C. S. Lim and A. Yamada, \MPL{7}, (1992), 2789.}
\nref\ma{E. Ma and P. Roy, \PRL{68}, (1992), 2879.}
\par
The standard model is working quite well, however, there are questions to be
answered. The nature of the Higgs sector is one of
the most mysterious problems. The symmetry-breaking mechanism is still
unknown.
The Higgs sector in the standard model is responsible
for breaking the $SU(2)\times U(1)$ symmetry and for making gauge-bosons and
fermions
massive. A lot of models, multi-Higgs-doublet models, technicolor models,
supersymmetric models
$\cdots$, have been proposed with their own properties in the
Higgs sector. None of new particles in these models, however, have been
detected
yet.
Recently, by using the oblique parameter $S$ and $T$ in addition to the
highly
accurate low-energy measurements,
it has become possible to impose some constraints on various kinds of models
beyond
the standard model. Originally, Peskin and Takeuchi estimated the
contribution
coming from QCD-like techicolor model to $S$ and $T$. They found that the
predictions
for $S$ in the model were outside experimentally allowed regions \pt.
Many authors performed estimations of the oblique parameters in various kinds
of models \refs{\dr{--}\ma}.
The oblique parameters are
useful and transparent tools for probing or constraining new physics beyond
the
standard model.
\nref\moor{C. D. Froggatt, R. G.
Moorhouse and I. G. Knowles, \PRD{45}, (1992), 2471.}
\nref\pv{A. Pomarol and R. Vega,
Preprint, SLAC-PUB-6196, SCIPP-93/08, SMU HEP 93-06.}
\par
In this paper we shall study the oblique parameters in electroweak theory
with two massless Higgs doublets. In a previous paper \ref\fkt{K. Funakubo,
A. Kakuto and K. Takenaga, Preprint, KYUSHU-HET-10, SAGA-HE-50.} we
investigated phase structure of this model by analyzing the effective
potential at
finite temperature. We found that the phase transition was of first order
and that the critical temperatures were significantly low compared with those
in
the standard model for wide ranges of scalar masses. We also estimated the
``critical'' temperatures for bubble nucleation and they were very close to
the
critical temperatures. We discussed that this model had many attractive
features for
electroweak baryogenesis even though there is no new CP violation.  In this
model
there are four kinds of scalar particles: charged Higgs
$H^{\pm}$, CP odd Higgs $A$, neutral Higgs (CP even) $h$ and scalon $H$.
Their masses are denoted by $M_{H^{\pm}}, M_A, M_h$ and $M_H$, respectively.
All these masses are free parameter except for $M_H$. We will find some
constraints on
these scalar masses by studying the oblique parameters $S$ and
$T$ in this model. This model has two Higgs doublets
which are massless by assumption. The oblique parameters in general
two-Higgs-doublet model were studied in \refs{\moor,\pv}
and the full expressions for them were obtained.
But numerical analyses for these parameters are very difficult because
there are six free parameters appearing in the expressions for
the oblique parameters \pv. Then one usually considers some special cases
such that there
are presumable mass hierarchies among new scalar particles. In the model with
two-massless-Higgs doublets the situation changes drastically. This is
because
symmetry breaking occurs by radiative corrections to yield a non-trivial
mass relation between the masses of scalon and other particles.
In order to break the
$SU(2)\times U(1)$ symmetry to
the $U(1)_{em}$, we have to impose flatness condition on the Higgs potential
along a suitable direction in VEV space of the Higgs fields \ref\gw{E.
Gildner and S.
Weinberg, \PRD{15}, (1976), 3333.}. The scalon, which is associated
with scale invariance of this model, becomes massive after symmetry
breaking. After rotating the Higgs doublets such that only one kind of
the Higgs field takes the VEV, the scalar mass terms in the Higgs potential
are automatically diagonalized. As a result there is no angle dependence in
the
oblique parameters.
In general two-Higgs-doublet model we need two different angles to
diagonalize the scalar mass terms, which are usually denoted by
$\alpha$ for neutral(CP even) sector and
$\beta$ for charged and neutral(CP odd) sector, so that the oblique
parameters
depend on these two angles \moor.
In our case these two angles are the same because
of the special direction in the VEV space and a flatness condition mentioned
above.
Thus the oblique parameters in our model is written in terms of only two free
parameters. The parameter $T$ has quadratic
dependences on the charged Higgs mass $M_{H^{\pm}}$ \refs{\moor,\pv}.
It is usually believed that
the constraint on $M_{H^{\pm}}$ from $T$ is strict due to its quadratic
dependence.
But we find that large mass differences between
$M_{H^{\pm}}$ and two scalar
masses $M_A, M_h$ could cancel the $M_{H^{\pm}}^2$ contribution to $T$.
Therefore
$M_{H^{\pm}}$ is not so restricted from $T$ as it was expected. On the other
hand the
parameter
$S$ can be both positive and negative. Heavy charged Higgs can shift $S$
to negative value but this is impossible when the neutral scalars are heavy
because of the mass relation in this model mentioned above.
We find that it is impossible to obtain a large negative $S$.
\par
In the next section we briefly review the Higgs sector
of electroweak theory with two massless Higgs doublets. In section 3 the full
expressions for the oblique parameters in our model are presented
and the behaviors of $S$ and $T$ are discussed briefly.
In section 4 we show the
results of numerical analyses of
$S$ and $T$ for two different reference points and discuss allowed regions
for
the scalar masses.

\newsec{The Higgs sector}
We work in the usual $SU(2)\times U(1)$ gauge theory of electroweak model
with
two massless Higgs doublets $\Phi_1,\Phi_2$ and with 3 generations of quarks
and
leptons. There exist dangerous tree-level flavour-changing neutral
current interactions mediated by scalar exchanges if we allow the most
general Yukawa
couplings. In order to avoid it, we restrict allowed Yukawa couplings by
imposing
a discrete symmetry\ref\wein{S. Weinberg, \PRL{37}, (1976), 657.}:
$\Phi_2 \rightarrow -\Phi_2, \quad u_{RA} \rightarrow -u_{RA}$.
Then we have the
following Yukawa couplings:
\eqn\ab{
\CL_Y = - \sum_{A,B}^{3}\big[{\bar Q_{LA}} {f_{AB}^{(u)}}{\tilde \Phi}_2
{u_{RB}}+
{\bar Q_{LA}} {f_{AB}^{(d)}} {\Phi_1 }{d_{RB}}+ {\bar l}_{LA}
f_{AB}^{(l)}\Phi_1
e_{RB} +{\rm h.c.} \big] ,} where ${\tilde\Phi}_2=i{\tau}^2 \Phi^*_2$ , with
$
\tau^a (a=1,2,3)$ being Pauli matrices. $Q_{LA}, l_{LA}$ are $SU(2)$ doublets
of quarks and leptons respectively. A subscript $A$ distinguishes
generations. The
discrete symmetry must also be imposed on the Higgs self-interactions,
otherwise the
Yukawa couplings \ab\  lose its meaning \ref\IKN{K. Inoue, A. Kakuto and Y.
Nakano, \PTP{63},(1980), 234.}. Then the Higgs potential is written as
\eqn\ac{\eqalign{ V_{H}&={\lambda_1 \over 2}({\Phi_1}^\dagger \Phi_1)^2+
{\lambda_2
\over 2}({\Phi_2}^\dagger \Phi_2)^2+
\lambda_3 ({\Phi_1}^\dagger \Phi_1)({\Phi_2}^\dagger \Phi_2)\cr &+\lambda_4
({\Phi_1}^\dagger \Phi_2)({\Phi_2}^\dagger \Phi_1)+ {\lambda_5 \over
2}[({\Phi_1}^\dagger \Phi_2)^2+{\rm h.c.}] \quad (\lambda_{1,2} >0).\cr } }
In our model the Higgs potential is a homogeneous polynomial of Higgs fields
so
that spontaneous symmetry breaking does not occur at the tree-level. We must
consider at least the one-loop potential. In the previous paper we discussed
how to obtain one-loop effective potential in this model. Calculating the
contribution from one-loop order to a potential in a model with more than one
Higgs
doublets is not so straightforward as it appears. Discussions are given in
\refs{\gw,\IKN}. Here we quote only results. The direction in the VEV space
of the Higgs
fields, which is consistent with the final $U(1)_{em}$ invariance, is
determined as\foot{We choose the phase convention of the Higgs field so that
the
sign of $\lambda_5$ is negative.}
\eqn\ad{
\vev{\Phi_1}={\rho \over {\sqrt 2}}{0 \choose n_{01}}, \quad
\vev{\Phi_2}={\rho \over {\sqrt 2}}{0 \choose n_{02}}, }
where $n_{01}$ and $n_{02}$ are defined by
$$ n_{01}^2\equiv {\sqrt \lambda_2 \over{\sqrt \lambda_1 + \sqrt
\lambda_2}},\quad n_{02}^2\equiv {\sqrt \lambda_1 \over{\sqrt \lambda_1 +
\sqrt
\lambda_2}}.
$$
The parameter $\rho$ is fixed by $v^2=1/{{\sqrt 2}G_F}$ after spontaneous
symmetry
breaking. Here $G_F$ is the Fermi coupling constant. The flatness condition
on the
Higgs potential is satisfied if
\eqn\ah{
\lambda \equiv {\sqrt{\lambda_1 \lambda_2}}+\lambda_3+\lambda_4+\lambda_5=0.
}
The condition \ah\ guarantees the minimum value of the tree-level
potential to be zero. Now we perform a rotation of the Higgs fields such that
only one Higgs field takes VEV:
\eqn\ai{
{\ph_1 \choose \ph_2} \equiv \left(
\matrix{C_{\beta}& -S_{\beta} \cr  S_{\beta} & C_{\beta}\cr } \right){\Phi_1
\choose \Phi_2}
.}
Here $\beta$ is defined by
$$
\tan \beta \equiv{{\rho n_{01}}\over{\rho n_{02}}}
\equiv {v_1 \over v_2}, \qquad {\rm with}\quad v_1^2+v_2^2=v^2.
$$
$C_{\beta}$ and $S_{\beta}$ are abbreviations for $\cos\beta$ and $\sin\beta$
respectively.
It is clear that $\ph_2$ plays the same role of the Higgs in the standard
model and
$\ph_1$ represents an independent extra doublet.
We expand $\ph_1$ and $\ph_2$ around the VEV as follows:
\eqn\aj{
\ph_1 ={ H^+ \choose {{1\over \sqrt 2}(h+iA)}},\qquad
\ph_2 ={G^+ \choose {{1 \over \sqrt 2}(v+H+iG^0)}}.}
Inserting \aj\ into the Higgs potential, we find that the scalar mass terms
are
given by
\eqn\ak{\eqalign{
V_H^{mass}&=
\big(H^-, G^- \big)\left(\matrix{ M_{H^{\pm}}^2 & 0 \cr
0 & \xi g^2 v^2/4 \cr} \right){H^+ \choose G^+}\cr &+
{1 \over 2}\big(A, G^0 \big)\left(\matrix{ M_A^2 & 0 \cr
0 & \xi(g^2+g'^2) v^2/4 \cr} \right){A \choose G^0}\cr &+
{1 \over 2}\big(h, H \big)\left(\matrix{ M_h^2 & 0 \cr
0 & 0 \cr} \right){h \choose H}\cr},
}
where $g$, $g'$ and $\xi$ are $SU(2),U(1)$ and gauge-parameter respectively.
The scalar masses in \ak\ are given as
$$
M_{H^{\pm}}^2(v) \equiv{1\over
2}({\sqrt{\lambda_1\lambda_2}}+\lambda_3)v^2,\quad
M_A^2(v)  \equiv -\lambda_5 v^2,\quad
M_h\equiv {\sqrt {\lambda_1 \lambda_2}}v^2.
$$
The scalon denoted by $H$ becomes massive after spontaneous symmetry
breaking. We identify the scalon as the usual Higgs boson in the standard
model. Its mass is defined by the
inverse propagator evaluated at zero momentum and is given by
\eqn\al{ M_H^2 =\biggl[{d^2 V_1 \over d\rho^2}\biggr]_{\rho=v}
     ={G_F \over {4 {\sqrt 2}\pi^2}}\biggl[6M_W^4+3M_Z^4-12M_t^4
                      +2M_{H^{\pm}}^4+M_h^4+M_A^4 \biggr].
}
Here we have ignored fermion masses except for the top quark mass. $V_1$ is
the
effective
potential of our model in one-loop approximation.
Detailed results are given in \refs{\fkt,\IKN}.

\newsec{The oblique parameters}
The oblique parameters $S$, $T$ and $U$ are defined in terms of gauge-boson
self-energies:
\eqn\ba{\eqalign{
\alpha_{em}S&\equiv 4e^2\big[\Pi^{\prime}_{33}(0)-\Pi^{\prime}_{3Q}(0)\big],
\cr
\alpha_{em}T&\equiv {e^2
\over{s^2c^2M_Z^2}}\big[\Pi_{11}(0)-\Pi_{33}(0)\big], \cr
\alpha_{em}U&\equiv 4e^2\big[\Pi^{\prime}_{11}(0)-\Pi^{\prime}_{33}(0)\big],
\cr
}}
where we adopt the notation of Peskin and Takeuchi \pt, and $e$ is the
electromagnetic
charge, $\alpha_{em}$ is the electromagnetic fine structure constant, and
$s^2\equiv
\sin^2\theta_W, c^2\equiv \cos^2\theta_W$. The indices $1 \sim 3$ in the
vacuum
polarization amplitudes refer to $SU(2)_L$ currents, while $Q$ refers to the
$U_{em}(1)$ current. Note that the $S$, $T$ and $U$ do not include the
standard
model contributions. The oblique parameters defined above describe the
effects of radiative corrections from new physics beyond the standard model.
In our model the new physics contributions to \ba\ are isolated from those of
the
standard model. Only new scalar particles $M_{H^{\pm}}, M_A$ and $M_h$
contribute to \ba.
It is straightforward to calculate the contributions
to \ba. The results are
\eqn\bb{\eqalign{
S&={1\over {12\pi}}\big[ g(M_A^2,M_h^2)+{\rm ln}{{M_A M_h} \over
M_{H^{\pm}}^2}
\big],\cr
T&={1 \over {16\pi}}{1 \over {s^2 c^2 M_Z^2}}\big[f(M_A^2,M_{H^{\pm}}^2)+
f( M_h^2,M_{H^{\pm}}^2)-f(M_A^2,M_h^2)\big], \cr
U&={1 \over {12\pi}}\big[g(M_A^2,M_{H^{\pm}}^2)+
g(M_h^2,M_{H^{\pm}}^2)-g(M_A^2,M_h^2)\big],\cr}
}
where we have introduced functions $f$ and $g$ defined by
\eqn\bc{\eqalign{
f(a,b)&\equiv {{a+b}\over 2}-{{ab}\over{a-b}}{\rm ln}{a\over b} \ge 0,\quad
(a,b>0),
\cr g(a,b)&\equiv -{5\over 6}+{{2ab}\over{(a-b)}^2}+
{{(a+b)(a^2-4ab+b^2)}\over{2{(a-b)}^3}}{\rm ln}{a \over b},\quad (a,b>0).\cr}
}
Note that $f$ and $g$ vanish for $a=b$ because both functions are
proportional to
square of mass difference $(a-b)^2$ when $|a-b|\ll a,b$. We see from \bb\
that $T$
has quadratic dependence on the charged Higgs mass $M_{H^{\pm}}$:
\eqn\bc{\eqalign{
T={1 \over {16\pi}}{1 \over {s^2 c^2 M_Z^2}}\biggl[M_{H^{\pm}}^2-
{{M_A^2 M_{H^{\pm}}^2}\over {M_A^2-M_{H^{\pm}}^2}}{\rm ln}
{M_A^2 \over M_{H^{\pm}}^2} &-
{{M_h^2 M_{H^{\pm}}^2}\over {M_h^2-M_{H^{\pm}}^2}}{\rm ln}
{M_h^2 \over M_{H^{\pm}}^2}\cr &+
{{M_A^2 M_h^2}\over {M_A^2-M_h^2}}{\rm ln} {M_A^2 \over M_h^2} \biggr].\cr
}} \par
The second and the third terms in square brackets in \bc\ are important for
smearing
the leading quadratic mass dependence when
mass differences between $M_A$ and $M_{H^{\pm}}$ or $M_h$ and $M_{H^{\pm}}$
become
large. This is because when $a\gg b$, the second term in the function $f$
behaves as
\eqn\bd{
{ab\over {a-b}}{\rm ln}{a \over b}\simeq b\times{\rm ln}{a \over b}.
}
The right hand side in \bd\ becomes large enough if $b$ takes appropriate
large
value. So the second and the third terms could drive $T$ to negative value
and smear
out the leading quadratic dependence on $M_{H^{\pm}}$. The value of
$M_{H^{\pm}}$
in this case must be relatively small, otherwise the term $M_{H^{\pm}}$ would
become
dominant contribution to $T$. In our model, however, relatively small charged
Higgs mass can be obtained automatically when the neutral Higgs scalars
become
heavy. From \al\ we see that the charged Higgs boson $M_{H^{\pm}}^2$ is
essentially
written in terms of two neutral scalars $M_A, M_h$\foot{This is because we
fix
$M_t$ and $M_H$ as the reference values for the standard model.}:
\eqn\be{
M_{H^{\pm}}^2=\sqrt{\big(
{{2{\sqrt 2}\pi^2}\over G_F}M_H^2-{1 \over
2}(6M_W^4+3M_Z^4-12M_t^4)\big)-{1 \over 2}(M_A^4+M_h^4)}.
}
Then as $M_A, M_h$ become heavier, $M_{H^{\pm}}$ becomes lighter.
Moreover, inserting \be\ into $T$ we can reduce free parameters
to two, {\it i.e.}, $M_A$ and $M_h$. So it is easy to perform numerical
analyses in
terms of two free parameters and one can simultaneously obtain the charged
Higgs mass
by \be\ each
time $M_h,M_A$ are fixed. The fourth term in a square bracket contributes to
$T$ additively when mass difference $M_A-M_h$ is large.\par
On the other hand, the function $g$ in $S$ increases monotonously
as mass differences become large. The terms that could be negative in $S$ is
the
second term only in square brackets. In order to make this
term negative, the argument in the logarithm must be smaller than one.
For small value of the neutral scalar masses, this is possible by \be. But it
is
impossible for heavy neutral scalar masses. The contributions from new scalar
particles to $S$ are all
logarithmic so that a large changes of the value in $S$ can not be expected.
$S$ also depends on only two free parameters by \be\ as in $T$.
\newsec{Numerical results and conclusions}
\nref\bb{G. Bhattacharyya, S. Banerjee and P. Roy, \PRD{45}, (1992), R729.}
\nref\ren{P. Renton, Z. Phys. C56, (1992), 355.}
It is necessary for us to make clear the reference point to define the
parameters
$S$ and $T$.
We choose two different reference points:
${\rm  I} (M_t,M_H)=(140,100){\rm GeV},  {\rm  II} (M_t,M_H)=(140,300){\rm
GeV}$.
At these reference points the experimental limits on $S$ and $T$ are given
by \refs{\bb,\ren}
\eqn\cb{\eqalign{
S=-0.76\pm 0.71, \qquad &T=-0.70\pm 0.49 \quad {\rm for\quad I}, \cr
S=-1.03\pm 0.66, \qquad &T=-0.46\pm 0.41 \quad {\rm for\quad II}. \cr}
}
The cesium data on atomic parity violation is included in both cases.
We see that $S$ and $T$ are both negative.
In analysis we
assume $M_A\ge M_h$ without loss of generality because $f$ and $g$ are
symmetric under $M_A \leftrightarrow M_h$.
In order to make the formulae
\ba\ valid, the scale of new physics must be larger than $M_Z$. We set the
minimum value of $M_h$ and $M_A$ to be $140$ GeV for illustration.
Perturbative triviality bounds on scalar masses in two-Higgs-doublet model
are
discussed in \ref\kc{D. Kominis and S. Chivukula, \PL{304}, (1993), 152.},
and
are given as
$M_{H^{\pm}}, M_A, M_h < 650
\sim 700$ GeV. We follow the results and set the upper bounds on the scalar
masses
to be
$650$ GeV.     \par
First, let us consider the case {\rm I}. In this case
the Higgs boson is relatively light so that it is natural to expect that
other
scalars are not extremely heavy. Actually, it is seen from \be\ that
extremely
heavy neutral scalars give the imaginary charged Higgs mass. The maximum
values
of $M_A$ and $M_h$, that make the charged Higgs mass real, are
$M_A=M_h \simeq 401$ GeV. So we assume that the upper bound on $M_A$ is $450$
GeV\foot
{In terms of quartic coupling constant, this value corresponds to
$|\lambda_5|/{4 \pi} \simeq 0.25$.}. This gives us a perturbative bound on
$M_A$.
In \fig\lht{The allowed mass regions for $M_{H^{\pm}}$ vs. $M_h$ from $T$ at
{\rm
I}.  The perturbative bound on $M_A=450$ GeV is depicted.}
we display the allowed mass regions for $M_{H^{\pm}}$ and $M_h$.
\vskip2pt
\centerline{{\bf Fig. 1}}
\vskip2pt\noindent
The minimum and
maximum values for $M_h, M_A$ and $M_{H^{\pm}}$ in the allowed regions are
$$
{\eqalign{
         M_h(min) &=140 {\rm {GeV}},            \cr
         M_A(min) &=369 {\rm {GeV}},            \cr
 M_{H^{\pm}}(min) &\simeq 223 (252) {\rm GeV}, \cr  }
\quad \eqalign{
          M_h(max) &=291 {\rm {GeV}},       \cr
          M_A(max) &=461 (450) {\rm {GeV}}, \cr
 M_{H^{\pm}}(max)  &\simeq 358 {\rm GeV}.   \cr}
}
$$
Large mass regions for $M_{H^{\pm}}$ are allowed even though
$T$ has the quadratic dependence on $M_{H^{\pm}}$. The charged Higgs boson
mass is
not so restricted as it was expected.
The value in parenthesis corresponds to the case where we put the
perturbative bound
on $M_A$.
In \fig\lhs{The allowed mass regions for $M_{H^{\pm}}$ vs. $M_h$ from $S$ at
{\rm
I}.} we show the allowed scalar mass regions from the experimental limit on
$S$
at the reference point {\rm I}.
\vskip2pt
\centerline{{\bf Fig. 2}}
\vskip2pt \noindent
{}From fig. 2 we see that the allowed mass
regions for $M_A$ and $M_h$ are very narrow and located in light mass
regions. This is
because if neutral scalars are heavy, then the charged Higgs boson is light
by
\be. So in this case there are no terms that give a negative value in $S$. We
find
that
$$
{\eqalign{
M_h(min)&=140 {\rm {GeV}}, \cr M_A(min)&=140 {\rm {GeV}}, \cr} \quad
\eqalign{
M_h(max)&=155 {\rm {GeV}}, \cr
 M_A(max)&=170 {\rm {GeV}}. \cr}
}
$$
The mass of the charged Higgs boson can not vary so much and its allowed mass
region is almost a point: $M_{H^{\pm}} \simeq 400$ GeV. If we combine the
constrains on the scalar masses from both $S$ and $T$, we see that
there is no allowed  mass regions of the scalar masses. The experimental
limit on
$S$ is very strict in this case if we believe the face value of $S$.
But the cesium data on atomic parity  violation, which is significant for
$S$, is
still ambiguous from both experimental and theoretical points of view.
The contribution from atomic parity violation to $S$ is very sensitive and
its data
gives $S_{Q_W}=-2.7\pm 2.0\pm1.1$\mr\ by itself. Reducing the total
uncertainties
in $S$ is very important as mentioned in \mr\ not only for our model but also
for other models. If we convert the allowed mass regions from
$T$ into the values of $S$ we have
\eqn\cc{
-1.50\times 10^{-2} \le S \le 2.40\times 10^{-2}.
}
$S$ is bounded above by a small positive value. This positive value is easily
obtained due to the uncertainties in $S_{Q_W}$.
At this stage the constrains on scalar masses from $T$ are more reliable than
those
from $S$.
In
\fig\mxt{$T_{min(max)}$ vs. $M_h$ at {\rm I}.
The area between the two lines $T_{min}, T_{max}$ are allowed.
The perturbative
bound on $M_A$ is set to be $450$ GeV. The calculation is stopped at
$M_h=M_A=401$
GeV to avoid the imaginary charged Higgs boson mass.} and \fig\mxs{
$S_{min(max)}$ vs. $M_h$
at {\rm I}.  The area between the two lines $S_{min}, S_{max}$ are allowed.
The perturbative
bound on $M_A$ is set to be $450$ GeV. The calculation is stopped at
$M_h=M_A=401$
GeV to avoid the imaginary charged Higgs boson mass.}
 we show the changes of the values in $T_{min(max)}, S_{min(max)}$ against
$M_h$ in order to clarify the possible values of $S$ and $T$ in our model.
\vskip2pt
\centerline{{\bf Fig. 3},\quad {\bf Fig. 4}}
\vskip2pt\noindent
In this calculation the perturbative bound on
$M_A$ is assumed and we stop the calculation at $M_A=M_h=401$ GeV in order to
avoid
the imaginary charged Higgs mass.\par
The same analyses are performed at the reference point {\rm II}.
As in the reference point {\rm I} we find the allowed mass regions for
$M_{H^{\pm}},M_h$ from $T$ which are displayed in
\fig\hht{The allowed mass regions for $M_{H^{\pm}}$ vs. $M_h$ from $T$ at
{\rm
II}.}.
\vskip2pt
\centerline{{\bf Fig. 5}}
\vskip2pt\noindent
In this calculation we set the triviality bounds on the
scalar masses discussed below eq. \cb. We obtain
$$
{\eqalign{
M_h(min)&=  140 {\rm  GeV}, \cr
M_A(min)&=  593 {\rm  GeV}, \cr
M_{H^{\pm}}(min)&\simeq  544 {\rm  GeV}, \cr}
\quad \eqalign{
M_h(max)&=  554 {\rm  GeV}, \cr
M_A(max)&= 650 {\rm  GeV},\cr
M_{H^{\pm}}(max)&\simeq 616 {\rm  GeV}.\cr}
}
$$
As for $S$ at this reference value there is no allowed mass regions that
explain
the experimental limit on $S$.
If we convert the allowed mass regions from
$T$ into the value of $S$ we have
\eqn\cd{ -1.97\times 10^{-2} \le  S \le  0.47\times 10^{-2}.}
This upper bound on $S$ is obtainable due to the uncertainties in atomic
parity violation as in the case of {\rm I}. Fig. 6\nfig\mxht{ $T_{min(max)}$
vs.
$M_h$ at {\rm II}.
The area between the two lines $T_{min}, T_{max}$ are allowed.
The calculation is stopped at the triviality bound $M_h=M_A=650$
GeV.} and fig. 7\nfig\mxhs{ $S_{min(max)}$  vs. $M_h$ at {\rm II}.
The area between the two lines $S_{min}, S_{max}$ are allowed.
The calculation is stopped at the
triviality bound $M_h=M_A=650$ GeV.}  correspond to fig. 3 and fig. 4 in
different
reference point.
\vskip2pt
\centerline{{\bf Fig. 6}, \quad {\bf Fig. 7}}
\vskip2pt\noindent
\par
We have not assumed any mass hierarchies among the scalar particles.
If we assume $M_{H^{\pm}}=M_A$ the custodial symmetry exists in the Higgs
potential\foot{$M_{H^{\pm}}=M_h$ also corresponds to the custodial symmetry
in the Higgs potential. The differences between them correspond to the
assignment
for CP eigenstate for each scalar particles \pv.}. In this case the
contribution to
$T$ is zero because of the property of the function $f$. On
the other hand,
$S$ has the non-trivial contributions from the scalar particles \ILY.
The changes of the values in $S$ are shown in \fig\suct{The behavior of
$S_{custodial}$ vs. $M_h$ at {\rm I}.},
 \fig\cuss{The behavior of $S_{custodial}$ vs. $M_h$ at {\rm II}.}
for the reference values {\rm I}, {\rm II} respectively.
\vskip2pt
\centerline{{\bf Fig. 8},\quad {\bf Fig. 9}}
\vskip2pt\noindent
The possible values of $S$ at the reference point {\rm I} for $M_h=140\sim
450$ GeV
are
$$\eqalign{
-1.64\times 10^{-2} &\le S_{custodial} \le 1.97 \times 10^{-2},  \cr
 247{\rm GeV}&\le M_{H^{\pm}}=M_A \le 362 {\rm GeV}.
}
$$
At the reference point {\rm II} we have
$$\eqalign{
-1.98\times 10^{-2} &\le S_{custodial} \le 0.52 \times 10^{-2},  \cr
541{\rm GeV}&\le M_{H^{\pm}}=M_A \le 616 {\rm GeV} \cr
}
$$
for $M_h=140\sim 650$ GeV. Both positive and negative value are possible.
The behavior of $S_{custodial}$ is similar with those of $S$ without the
custodial
symmetry so that it does not take large negative values.   \par
We have studied the oblique parameters $S$ and $T$ in electroweak theory with
two
massless Higgs doublets at the two different reference points.
We have performed numerical analyses of $S$ and $T$ by
using the specific properties in this model. We have found that the effects
of quadratic dependence on the charged Higgs mass in $T$ could be cancelled
and the
constraint on $M_{H^{\pm}}$ from the experimental limit on $T$ is not so
stringent
for both {\rm I} and {\rm II} as it was expected. On the other hand, the
experimental limit on $S$ for both  {\rm I} and {\rm II} are very severe for
our model
if we believe the face values \cb. We can not say that there are no allowed
mass
regions for the scalar particles that satisfy the experimental limits on both
$S$ and $T$ in this model because large uncertainties from atomic parity
violation exist in $S$. It is interesting that the future experiments reduce
the
uncertainties to the accuracy such that we can conclude  whether our model is
excluded or not.
%


\bigbreak\bigskip
\centerline{{\bf Acknowledgments}}\nobreak
The author would like to thank A.Kakuto for careful reading of the
manuscript.
He also thanks K. Inoue and T. Inami for valuable discussions.




\listrefs
\listfigs

\bye